# Impact of Small Phonon Energies on the Charge-Carrier Lifetimes in Metal-Halide Perovskites

Thomas Kirchartz[1,2,*], Tom Markvart[3,4], Uwe Rau[1], David A. Egger[5,*]

[1]IEK5-Photovoltaics, Forschungszentrum Jülich, 52425 Jülich, Germany
[2]Faculty of Engineering and CENIDE, University of Duisburg-Essen, Carl-Benz-Str. 199, 47057 Duisburg, Germany
[3]Centre for Advanced Photovoltaics, Czech Technical University, 166 36 Prague 6, Czech Republic
[4]Engineering Sciences, University of Southampton, Southampton SO17 1BJ, UK
[5]Institute of Theoretical Physics, University of Regensburg, 93040 Regensburg, Germany

**Abstract**

Solar cells based on metal-halide perovskite absorber layers have resulted in outstanding photovoltaic devices with long non-radiative lifetimes as a crucial feature enabling high efficiencies. Long non-radiative lifetimes occur if the transfer of the energy of the electron-hole pair into vibrational energy is slow, due to, *e.g.*, a low density of defects, weak electron phonon coupling or the release of a large number of phonons needed for a single transition. Here, we discuss the implications of the known material properties of metal-halide perovskites (such as permittivities, phonon energies and effective masses) and combine those with basic models for electron-phonon coupling and multiphonon-transition rates in polar semiconductors. We find that the low phonon energies of MAPbI$_3$ lead to a strong dependence of recombination rates on trap position, which can be readily deduced from the underlying physical effects determining non-radiative transitions. Here, we show that this is important for the non-radiative recombination dynamics of metal-halide perovskites, as it implies that these systems are rather insensitive to defects that are not at midgap energy. This can lead to long lifetimes, which indicates that the low phonon energies are likely an important factor for the high performance of optoelectronic devices with metal halide perovskites.

**TOC Figure**

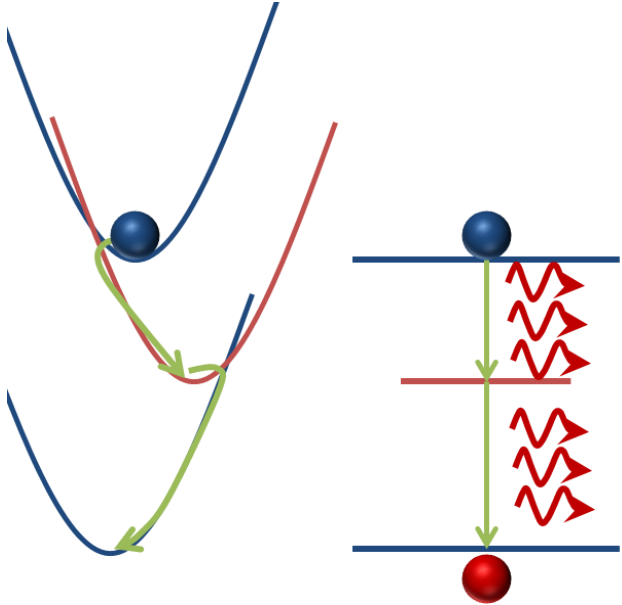

The high open-circuit voltages[1-5] and long non-radiative charge-carrier lifetimes[6-11] are key to understanding the exceptional[3,12-15] photovoltaic performance that has been reached using metal-halide perovskite (MHP) absorber materials. Understanding the reasons for these long lifetimes is of crucial importance because it would help develop the perovskite-based solar cell technology further by directing the discovery of alternative MHP compounds. These may have higher stability[1,3,16,17] than the original methyl-ammonium lead halide perovskites, while retaining the properties that enable long charge-carrier lifetimes. In addition, a clear identification of what microscopic properties of a material are beneficial for long lifetimes would also enable us to screen materials computationally for these desired properties.[18-20]

Among the early proposed explanations for the long lifetimes in MHPs was the observation that in density functional theory calculations most intrinsic defects (vacancy, interstitial, and substitutional defects) are either rather shallow, or have a high formation energy, and are thus unlikely to form.[18,21] Since then, there have been various additional suggestions attempting to explain how peculiar properties of lead-halide perovskites could lead to the long lifetimes. Among these suggestions is the existence of an indirect band gap[22-25] just below the direct one, the low density of states at the valence-band maximum,[26] and the formation of large polarons[27,28] that were thought to be screened from each other, thereby slowing down recombination. Rather than adding a new explanation, we here intend to estimate the charge-carrier lifetimes (for a given trap density) of MHPs based on existing theories[29-33] for *non-radiative recombination via the emission of multiple phonons*.

In general, we consider that recombination of electron-hole pairs in semiconductors may proceed either in a radiative process *via* the emission of photons, or in a non-radiative process *via* the emission of phonons or by the Auger effect which also leads to phonon emission eventually. Radiative and Auger-driven recombination are intrinsic mechanisms that occur also in a perfect, defect free single crystal.[34,35] Band-to-band recombination of an electron-hole pair *via* the emission of multiple phonons, however, is typically considered impossible in inorganic semiconductors. Thus, in most thin films of polycrystalline semiconductors the interruption of the crystal periodicity at grain boundaries, or the presence of intrinsic/extrinsic point defects, leads to the creation of localized states in the band gap. The existence of these states is typically thought to accelerate recombination by reducing the number of phonons that have to be emitted in a single transition. Similarly, a commonly stated but infrequently explained rule of semiconductor physics is that this acceleration of recombination events is strongest if the energy of localized states is situated around midgap. Recombination *via* such (singly charged) localized states in the band gap is typically described by Shockley-Read-Hall (SRH)[36,37] statistics. While the SRH model explains why recombination at a certain voltage is most likely proceeding *via* localized states that are between the two quasi-Fermi levels,[38-40] it does not explain why recombination *via* midgap energy states would be any different from recombination *via* other defect levels in between the two quasi-Fermi levels.[38] To better understand these phenomena and reveal their effect in MHP solar cells, we study the effect of electron-phonon coupling and the phonon energy, *i.e.*, the number of phonons required for a certain transition, on SRH lifetimes *via* multiphonon emission.

Coordination coordinate diagrams, as shown in Fig. 1, serve as the conceptual starting point for theories of multiphonon transitions in semiconductors or molecules. Let us assume, for instance, a transition of an electron in the conduction band to an available localized state in the band gap. These two states in general correspond to two different potential energy surfaces represented by the upper and lower parabola reflecting different configurations of the crystal (or the molecule) with minima shifted relative to each other on the coordination coordinate axis. The two states relevant for the transition we are considering are vibrational eigenstates of the lower and upper parabola. Recombination may now occur if the overlap between the lowest-energy wavefunction on the upper parabola overlaps with that of the highest-energy wavefunction on the lower one. Depending on the temperature and relative energetics, different paths from upper to lower parabola can be conceived, including thermally activated processes. Typically, the relative shift of the two parabolas relative to each other is described by the parameter $S_{HR}$, the Huang-Rhys factor. It is defined using the energy difference between the minimum of the upper parabola and its value at the minimum of the lower one, as indicated in Fig. 1a and b. If $S_{HR}$ is very small, *i.e.*, the two potential energy surfaces are hardly shifted relative to each other, the vibrational wavefunction overlap will be small. If, additionally, the transition energy approximated by the energy difference between the two minima of the parabolas, $\Delta E$, is large, the barrier for the recombination process will be large (Fig. 1a), which is known as the 'energy gap law'.[32,33,41,42] From this, it is immediately clear that for band-to-band recombination in high-quality inorganic semiconductors, due to the small associated Stokes shift and, hence, $S_{HR}$, band-to-band recombination *via* the emission of multiple phonons is highly unlikely to occur. Conversely, when the two parabolas are massively shifted (Fig. 1b) and the wavefunction overlap is large, the limit $S_{HR}E_{ph} = \Delta E$ is approached, where $E_{ph} = \hbar\omega$ is the energy for a phonon with frequency $\omega$. In this limit, the transition rate is maximized. Note that the Huang-Rhys factor is not a material constant, but depends on microscopic details of the states involved in the transition, and therefore varies between different scenarios.

Transition rate calculations according to Fig. 1 lead to a high temperature limit, in which the transition rate is thermally activated according to an activation energy, and a low temperature limit with a weak temperature dependence. Interestingly, this low temperature limit for the rate already shows a strong dependence on the number $p$ of phonons needed for a given transition ($p = \Delta E/E_{ph}$) and on the value of the Huang-Rhys factor $S_{HR}$. To see that this is a general physical effect, consider the multiphonon transition rate $k$ in a semi-classical approximation valid over the whole temperature range, which has been derived by Markvart[29]

$$k \propto \frac{\sqrt{2\pi}}{\hbar^2 \omega \sqrt{p\sqrt{1+x^2}}} \exp\left[ p\left( \frac{\hbar\omega}{2kT} + \sqrt{1+x^2} - x\cosh\left(\frac{\hbar\omega}{2kT}\right) - \ln\left(\frac{1+\sqrt{1+x^2}}{x}\right) \right) \right]. \quad (1)$$

Here, $T$ is temperature, $k$ the Boltzmann and $\hbar$ the reduced Planck constant. The parameter $x$ includes the dependence of the recombination rate on the ratio $p/S_{HR}$ and is defined as[29]

$$x = \begin{cases} \dfrac{S_{HR}}{p \sinh(\hbar\omega/2kT)} & \text{for } S_{HR} < p \\ \dfrac{p}{S_{HR} \sinh(\hbar\omega/2kT)} & \text{for } S_{HR} > p \end{cases}. \quad (2)$$

Following the derivations outlined by Ridley[43,44], it is possible to express the Huang-Rhys factor for polar coupling as

$$S = \frac{3}{2(\hbar\omega)^2} \left\{ \frac{q^2(M_r/V_0)\hbar\omega^2}{M_r \omega q_D^2} \left( \frac{1}{\varepsilon_\infty} - \frac{1}{\varepsilon} \right) \right\} I(-2, 2\mu, q_D a^* \nu/2) \quad (3)$$

where $M_r$ is the reduced mass, $q_D$ is the radius of a sphere with the Brillouin-zone volume, $\varepsilon_\infty$ and $\varepsilon$ are the high- and low-frequency limit of the dielectric function, and $a^*\nu/2$ is a rough estimate for the radius of the defect wavefunction in the 'quantum defect model'. The function $I$ is defined via

$$I(a,b,c) = \frac{1}{(bc)^2} \int_0^1 \frac{x^a \sin^2(b \tan^{-1}(cx))}{[1+(cx)^2]^b}. \quad (4)$$

From Eq. (2) it can be seen that the number of phonons and the value of $S_{HR}$ strongly influence the multiphonon transition rate. In fact, one finds that for $T \to 0$, $k \sim 1/p!$. Eqs. (3) and (4) use a series of parameters given in Table I for the calculations used to produce the data shown below. The energy of the defect level, *i.e.*, the trap depth, $\Delta E$, determines the function $I$, and thus $S_{HR}$, *via* the parameter $\nu = q(8\pi\varepsilon a^* \Delta E)^{-1/2}$. The charge state of the trap also enters the calculation of both through the parameter $\mu$: $\mu = \nu$ for positively charged defects, $\mu = 0$ for neutral defects, and $\mu = -\nu$ for negatively charged defects. Therefore, the quantum defect model we use here cannot account for all possibly important microscopic details and effects, but it does include the important dependencies of the transition rate on the energy and charge state of the defect.

For the prototypical MHP crystal CH$_3$NH$_3$PbI$_3$ (MAPbI$_3$) the phonon energies, the effective mass, and the reduced mass $M_r$ are known (see Table I and ref. [45]). In particular, the vibrational energies of MHPs such as MAPbI$_3$ have been reported from computations as well as infrared (IR) and Raman spectroscopy.[45-51] These studies have shown that the longitudinal optical (LO) phonons, corresponding to Pb-halide modes, are relatively low in energy, which is consistent with the mechanically soft lattice of MHPs.[52,53] We mention that in Eq. (3) it is assumed that the interaction is fully-screened *via* the "Pekar factor", defined as $\gamma = \varepsilon_\infty^{-1} - \varepsilon^{-1}$. However, it is *a priori* unclear which frequency range of $\varepsilon(f)$ is appropriate for describing polar coupling for a given transition. In absence of further information, we chose the experimentally-reported high- and low-frequency limit of $\varepsilon(f)$ (see Table 1), which maximizes polar coupling and, as discussed in detail below, counteracts the effect of low phonon energies we describe here. Therefore, we may now proceed to calculate $S_{HR}$ for polar coupling based on Eqs. (3) and (4).

Fig. 2a shows the result for MAPbI$_3$ as a function of the energy difference $\Delta E$ for the case of a positively charged, negatively charged, and neutral defect. $S_{HR}$ is quite high due to the above-mentioned low phonon energy, $E_{ph} = \hbar\omega = 16.5$ meV (LO phonon), and the high "Pekar factor", defined as $\gamma = \varepsilon_\infty^{-1} - \varepsilon^{-1}$. In addition, $S_{HR}$ increases for larger values of $\Delta E$ with the exception of very shallow negatively charged defects. Note that such large $S_{HR}$ values are not unprecedented[54] and not in conflict with recently reported, much smaller values for similar perovskite systems either,[55-57] as these experiments studied band-to-band transitions. We however are focussing on transitions between delocalized and localized states; the latter typically involve larger structural changes and larger values of $S_{HR}$. Also note that the radius of the defect wavefunction enters our computation via the parameter $a^*v/2$, as defined above.

Recombination *via* singly charged defects in semiconductors is typically described using Shockley-Read-Hall (SRH) statistics[36,37] with a recombination rate given by

$$R_{SRH} = \frac{np - n_0 p_0}{(n + n_1)\tau_p + (p + p_1)\tau_n}. \quad (5)$$

with the SRH lifetimes defined by $\tau_{n,p} = (k_{n,p} N_t)^{-1}$ and using the abbreviations $n_1 = N_C \exp[(E_T - E_C)/kT]$ and $p_1 = N_V \exp[(E_V - E_T)/kT]$. Here, $n$ is the electron concentration, $p$ the hole concentration, the index '0' represents the equilibrium concentrations, $N_T$ is the trap density, $N_C$ is the conduction band effective density of states, $N_V$ is the valence band effective density of states, $E_C$ is the conduction band edge, $E_V$ the valence band edge and $E_T$ the trap position. Using the prefactors for the recombination rate as given by Ridley and discussed in the supporting information (SI), we obtain an estimate of $k_n$ and $k_p$ based on Eq. (1). This estimate leads to rates that decrease strongly for higher energy differences between band and trap state. As illustrated in Fig. 2b, recombination *via* traps always involves two transitions, conduction band to trap and trap to valence band. Thus, max($E_C$-$E_T$, $E_T$-$E_V$) will dominate the rate and midgap traps will be most detrimental because max($E_C$-$E_T$, $E_T$-$E_V$) is minimized. For deep traps and forward bias, $np > n_0 p_0$, $n_1 \ll n$, and $p_1 \ll p$. Thus, the SRH recombination rate simplifies to $R_{SRH} \approx np/(n\tau_p + p\tau_n)$. If we additionally assume that we are in high-level injection (doping level is negligible relative to the optically or electrically injected charge carrier density), *i.e.*, $n = p$, we can write $R_{SRH} \approx n/(\tau_p + \tau_n) \equiv n/\tau_{eff}$. We know that effective lifetimes $\tau_{eff}$ of MAPI have been measured to be on the order of 1 μs.[8,58] Thus, we may now calculate the trap density needed to achieve a certain effective lifetime as $N_T \approx ((\tau_{eff} k_n)^{-1} + (\tau_{eff} k_p)^{-1})^{-1}$. We also include the effect of Coulomb attraction between positively charged trap states and electrons and negatively charged trap states and holes *via* the so-called Sommerfeld factors, $s_a$, as described in the SI.

The results for the case of a donor like and an acceptor like trap are shown in Fig. 2c as a function of the trap energy, $E_T$. Since the total rate will be limited by the slower of the two transitions, the resulting trap density required for a lifetime of 1 µs is symmetric with respect to midgap trap energies (~0.8 eV for MAPbI$_3$) and it increases sharply when the trap is energetically no longer at midgap. This is due to the strong dependence of the recombination rate, as approximately given by Eq. (1), on the number of phonons involved in the transition, in which a larger number of phonons leads to much slower rates. For MAPbI$_3$, due to the small optical phonon energy, a trap at midgap energy would require 49 phonons to be emitted instantaneously, making non-radiative recombination unlikely, which automatically results in the long lifetimes that were observed experimentally.

To put this finding into perspective, we mention that recently[41] the high energy intramolecular vibrations in organic molecules have been shown to be one main reason for high losses due to non-radiative recombination in organic solar cells. In particular, the unavoidable C=C double bond with a vibrational energy of ~160 meV, as present in the backbone of any conjugated polymer or small molecule, was identified as the main culprit. The key observation of ref [41] was that non-radiative voltage losses in organic solar cells follow the so-called energy gap law,[32,33,42] which states that the recombination rate decreases with $\Delta E$. The energy gap law can be assessed from Eq. (1) and, given that $\Delta E$ does not directly enter Eq. (1) but instead the ratio $p = \Delta E/E_{ph}$, one might conclude that lower phonon energies $E_{ph}$ would be key to achieving low non-radiative voltage losses. As shown in the SI, the product of phonon energy and reduced ionic masses only varies mildly over a wide range of relevant semiconductors. Therefore, the beneficial effect of low phonon energies would imply that semiconductors made of heavier atoms show advantages in terms of a lower likelihood of non-radiative recombination *via* multiphonon emission. We note that this expectation is reminiscent of vibrational transitions in molecules, where it was shown that deuterides, due to the heavier mass of the atomic oscillator, relax much slower than corresponding hydrides.[33,59,60]

To calculate the effect of this on non-radiative lifetimes in MHPs, we first note that the phonon energy also enters the equation for the Huang-Rhys factor (Eq. (3)), with higher phonon energies leading to lower Huang-Rhys factors. Thus, the effects on $p$ and $S_{HR}$ are partly cancelling in the evaluation of the rate (Eq. (1)). In the following, we keep the parameters for MAPbI$_3$ (see Table I) constant, and vary the reduced mass and phonon energy such that their product remains constant (see SI) to then recalculate $S_{HR}$ and the transition rates. Fig. 3a shows that, as expected, $S_{HR}$ strongly depends on phonon energy (and reduced mass) and decreases with higher phonon energies. In analogy to Fig. 2b, Fig. 3b shows the defect densities required for obtaining the generally accepted SRH lifetime of ~1 µs for MAPbI$_3$, when varying the phonon energy (and reduced mass) of the atomic oscillator. For higher phonon energies, the necessary defect density at midgap energies for obtaining the experimentally determined lifetime already decreases, *i.e.* lower phonon energies make multiphonon emission less likely.

Remarkably, however, the strongest effect is observed for defect levels away from midgap energy, which are the most relevant energies for practical MHP cells. Figure 3b shows that the necessary defect density to achieve a certain lifetime is much less sensitive to the defect energy position at high phonon energies than it is for lower ones. In the extreme situation of a large phonon energy of 160 meV, which mimics organic semiconductors, the position of the defect level becomes even somewhat unimportant, as non-radiative band-to-band transitions, which are very unlikely in inorganic semiconductors, become possible. This finding implies that low-energy phonons, for which the required defect density away from the midgap energy strongly increases, help suppress non-radiative recombination *via* harmful defects deep in the gap. This is also in line with the observation that ideality factors ~1 are common in fully depleted organic solar cells, a situation that is typically not encountered in fully depleted inorganic solar cells.[39,61] It may also explain why harmful defects[62-65] and higher ideality factors, close to two, are sometimes observed[66-68] in organic solar cells without massively deteriorating photovoltaic performance.

Thus, we can readily conclude that for recombination to be fast in a polar semiconductor with low phonon energies, not only the defect density must be high but, in addition, the position of the defect must be very close to midgap. Shifts of the state energy relative to the midgap energy will increase the number of phonons needed for the larger energy of the two transitions in the scenario discussed above, such that the total recombination rate will decrease substantially. This effect is much less pronounced for higher phonon energies as is the case in typical inorganic semiconductors, in which a smaller number of phonons is required for a single transition. Based on the multiphonon transition scenario, our data and analysis imply that a low density of deep defects (but, potentially, a high density of shallow defects) in a *material with low phonon energies leads to low non-radiative recombination rates and, thus, long carrier lifetimes*. In view of recent proposals for defect tolerance and healing in MHPs,[69,70] one may adapt the view that these phenomena are driven in part by the low-energy phonons of MHPs, which strongly reduce the detrimental impact of deep defects and associated non-radiative recombination. From this, the low phonon energies of MHPs emerge as an important 'microscopic materials design' parameter for further optimizing perovskite solar cells.

We consider this result as an important step towards establishing a deeper understanding of how recombination kinetics in addition to defect densities contribute to long charge-carrier lifetimes in perovskite solar cells. However, we would like to point out that models for multiphonon processes in semiconductors have a high level of complexity. In particular, the series of exponential terms in Eq. (1) leads to large changes in the result for small deviations in input parameters. In addition, strictly speaking the equations are still only approximations to an even more complex physical reality. For example, theoretical calculations showed that dephasing times in MPHs depend on their ionic composition and doping,[71] which influences electronic transitions, but is not considered in our model. In the following, we will therefore briefly discuss further potential shortcomings of the model as well as further steps for model development and experimental validation.

One of the key assumptions for the calculations presented in Figs. 2 and 3 is that polar coupling dominates multiphonon transitions. This assumption was made based on the high Pekar factor, $\gamma = \varepsilon_\infty^{-1} - \varepsilon^{-1}$, in MAPbI$_3$. For less polar semiconductors, deformation potential coupling due to optical phonons might be the dominant mechanism. For MAPbI$_3$, the effect of polar coupling, especially for different frequencies of the dielectric response, has not been systematically compared to the effect of deformation potential coupling due to optical phonons. While Ridley proposed an equation to also calculate $S_{HR}$ for deformation potential coupling, the actual calculation requires knowledge on the so-called optical deformation potential constant. This quantity is rather difficult to measure experimentally or calculate theoretically, and is currently unknown for MAPbI$_3$. Therefore, the deformation potential constant has in the past typically been used as an 'adjustable constant'.[72] In the SI, we therefore show the recombination rates for different values of the optical deformation potential constant. These data show that larger values of the optical deformation potential constant reduce the beneficial effect of low phonon energies on recombination, which means that our findings do not hold for all types of semiconductors. Within our model and parameters, we determine an upper limit for the optical deformation constant of $\sim 5 \times 10^8$ eV/cm for which a noticeable positive effect of the low phonon energy on recombination is still observed.

Furthermore, the theoretical approach discussed above makes very specific assumptions on the vibrations involved in multiphonon recombination: in MAPbI$_3$, high-energy vibrations are associated with the organic molecule. However, the electronic states of the organic cation are energetically far-off the frontier valence and conduction band states, which are determined by contributions from Pb and I atoms. Non-adiabatic molecular-dynamics computations have also shown that the Pb-halide stretching and bending modes couple strongly to the electronic structure, and are thus important factors determining the recombination rates at $MAPbI_3/TiO_2$ interfaces.[71,73] Therefore, in our single mode model we have chosen the highest-energy longitudinal optical phonon associated with Pb-I vibrations.[45] Thus, the model does not explicitly include information about multiple phonons, with individual frequencies and coupling constants, and their effect on multiphonon transitions, *i.e.*, the phonon density of states is disregarded. We note that a fully predictive model, including first-principles data of all the relevant modes and their coupling,[71,73] is far from trivial but conceivable in principle.[74,75] In practice, MHPs such as $MAPbI_3$ exhibit strong anharmonic vibrational contributions at room temperature,[45,50] complicated photo-excited carrier relaxation effects,[76,77] and intriguing decoherence effects on electronic transitions.[71] While these effects go far beyond the scope of our purely model-based approach, elucidating their impact on carrier lifetimes indeed motivates future studies. In the same spirit, a further general question about the validity of multiphonon transition theory is the inclusion of the effect of entropy.[78,79] For higher numbers of phonons needed for a given transition, the number of contributing possible combinations of phonon modes becomes increasingly larger. This effect gives rise to the Meyer-Neldel rule. The very general question how to merge theories of multiphonon recombination with entropy considerations is discussed in ref. [78] and currently still under debate.[79]

We also would like to put our finding that low phonon energies are key to reducing non-radiative recombination into perspective of other currently discussed effects explaining the lifetime of carriers in MHPs cells. First, several studies reported the impact of the indirect band gap[22,23,25,80] and polar organic molecule[81,82] on radiative and non-radiative lifetimes of MHPs. Interestingly, a recent terahertz spectroscopy study[83] on the all-inorganic MHP $CsPbI_3$ showed that systems lacking a polar cation can still exhibit similar bimolecular recombination rates as hybrid MHPs, which indicates that the polar cation does not strongly impact radiative transitions. The role of the polar organic cation is also discussed in the context of large polaron formation MHPs cells.[27,28] An interesting consequence of the theory of multiphonon recombination with polar coupling is that high Pekar factors, $\gamma = \varepsilon_\infty^{-1} - \varepsilon^{-1}$, in general lead to stronger non-radiative recombination, as they enhance $S_{HR}$ (see Eqs. (1)-(4)). This suggests that stronger polaronic effects, and higher values of $\gamma$, would not lead to slower band-to-band or defect-to-band recombination as is currently speculated. Note that this explanation is not questioning the formation of large polarons in MHPs *per se*. Rather, we here address the issue whether recombination is dictated by two (possibly screened) charges finding each other in a diffusion-limited recombination process, or by the time required to emit ~ 50 phonons. We tentatively conclude that the rate for multiphonon transition is the more relevant factor in limiting the lifetime, based on two observations: In organic photovoltaics, with materials that show orders of magnitude lower mobilities compared to MHPs, there is solid evidence for the effect of the energy gap law,[41] and hence phonon energies. This suggests that even for this case, Langevin-type recombination is not limiting the total recombination rate. In MHPs, Langevin-type recombination models indeed overestimate the recombination rates by orders of magnitude. From this, we conclude that the long lifetime in MHPs is likely not observed *because* of the high $\gamma = \varepsilon_\infty^{-1} - \varepsilon^{-1}$, *but despite* the high $\gamma = \varepsilon_\infty^{-1} - \varepsilon^{-1}$.

Finally, we comment on experimental determination of capture cross sections in perovskite solar cells as a way of testing our conclusions. In order to determine $S_{HR}$ for a specific defect-mediated transition one measures the temperature dependence of the capture cross section in the high temperature region of Eq. (1), where an activation energy becomes visible. This activation energy $E_B = (\Delta E - S_{HR}\hbar\omega)^2 / (4 S_{HR}\hbar\omega)$ solely depends on the trap depth, $S_{HR}$, and the phonon energy, which in principle allows for extraction of the Huang-Rhys factor if $S_{HR}$ is the only unknown. Last but far from least we mention that such studies would be interesting especially for MHPs, where the ionic mass can be changed by ionic composition, *e.g.*, by using Br instead of I, Sn instead of Pb, and different A-site cations. Such substitutions will change the ionic masses, thereby the optical phonon energies, which in light of our results is expected to impact recombination rates and lifetimes.

In summary, we have used model calculations to examine the effect of low-energy phonons on charge-carrier lifetimes of MHPs. Transitions between localized defect states and delocalized states were studied in a multiphonon transition scenario including the effect of defect energy and densities. We found that the long charge carrier lifetimes of $MAPbI_3$ can be explained in this scenario and showed that low-energy phonons are key towards achieving these lifetimes: In polar semiconductors with low-energy phonons, non-radiative recombination *via* deep traps is suppressed by many orders of magnitudes when compared to semiconductors with higher phonon energies, which can readily be deduced from the physics of multiphonon transitions. Here, we found that this effect becomes considerably more pronounced if the defect energy is some few 100 meV away from the midgap energy, *i.e.*, the impact of low phonon energies on non-radiative transitions is especially pronounced for the practically most relevant defect scenarios. Based on our findings, we conclude that even a relatively high density of deep defects is compatible with a low non-radiative recombination rate in a polar semiconductor with low phonon energies, especially if the defect energies are not directly at midgap. With this, the low phonon energies of MHPs emerge as an important microscopic material parameter that together with shallow trap energies could represent one important ingredient for the success of MHPs. This aspect needs to be taken into account when aiming to further advance perovskite solar cells.

Supporting Information

Explanation for the choice of the prefactor for the multiphonon transition rate. Sommerfeld factors to take Coulomb attraction into account. Temperature dependence of the rates. Dependence of reduced mass on phonon energy. Application of the methodology to $CH_3NH_3PbBr_3$ and comparison to $CH_3NH_3PbI_3$. Influence of deformation coupling.

Author Information

Corresponding Author:

[*]To whom correspondence should be addressed. E-Mail: t.kirchartz@fz-juelich.de, david.egger@physik.uni-regensburg.de

Acknowledgements

TK and UR acknowledge support from the DFG (Grant Nos. KI-1571/2-1 and RA 473/7-1). TM acknowledges support via the Centre for Advanced Photovoltaics supported by the Czech Ministry of Education, Youth and Sport (CZ.02.1.01/0.0/0.0/15_003/0000464). DAE acknowledges funding provided by the Alexander von Humboldt Foundation in the framework of the Sofja Kovalevskaja Award endowed by the German Federal Ministry of Education and Research. The authors would like to thank Vanessa Wood, Nuri Yazdani (both ETH Zürich), David Cahen, Omer Yaffe (both Weizmann Institute of Science), and Deniz Bozyigit (Battrion) for fruitful discussions.

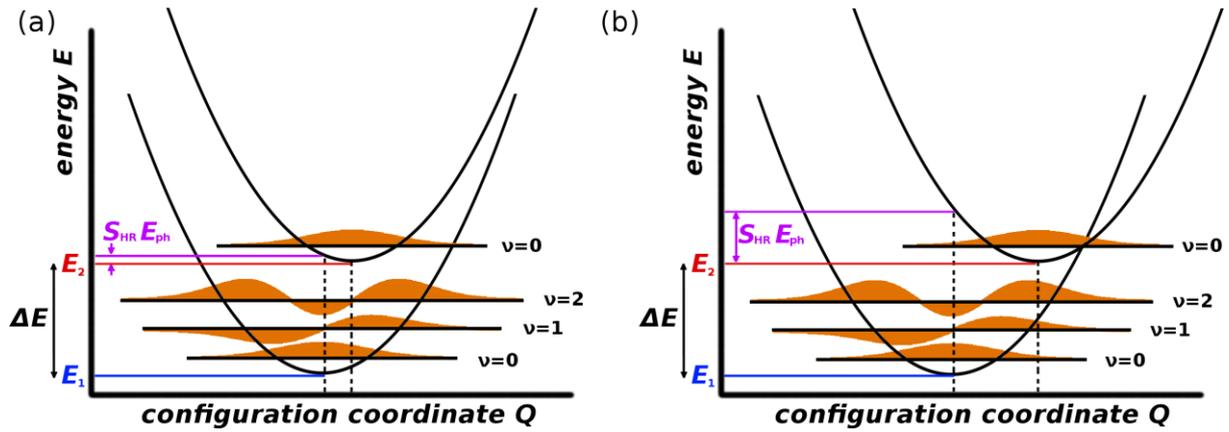

**Figure 1:** Schematic showing a configuration coordinate diagram with small (a) and large (b) Huang-Rhys factor, $S_{HR}$. $E_1$ and $E_2$ indicate the minimum energies of two representative potential energy surfaces that correspond to the two states in the multiphonon transition scenario, which are separated by an energy difference, $\Delta E$. $E_{ph}$ denotes the energy of the optical phonon emitted in the multiphonon process and ν is the vibrational quantum number.

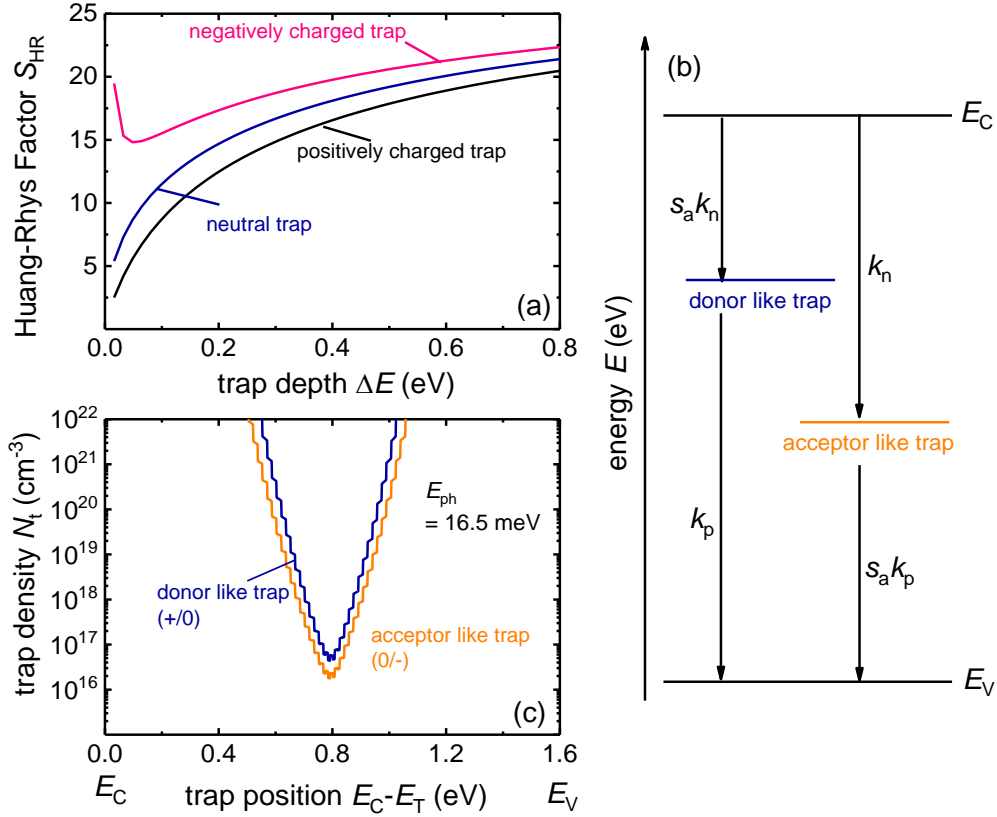

**Figure 2:** (a) Huang-Rhys factor as a function of the energy difference ΔE, as defined in Fig. 1, using the parameters given in Table I. The result changes with the charge state of the trap according to Eqs. (3) and (4). (b) Schematic of the transition from the conduction band to a donor and acceptor like trap, and from a donor and an acceptor like trap to the valence band. The respective rates, *k*, and the Sommerfeld factor, $s_a$, are indicated (see text for details). (c) Trap density needed to achieve a SRH lifetime of 1 μs for a phonon energy of 16.5 meV as given by ref. [45].

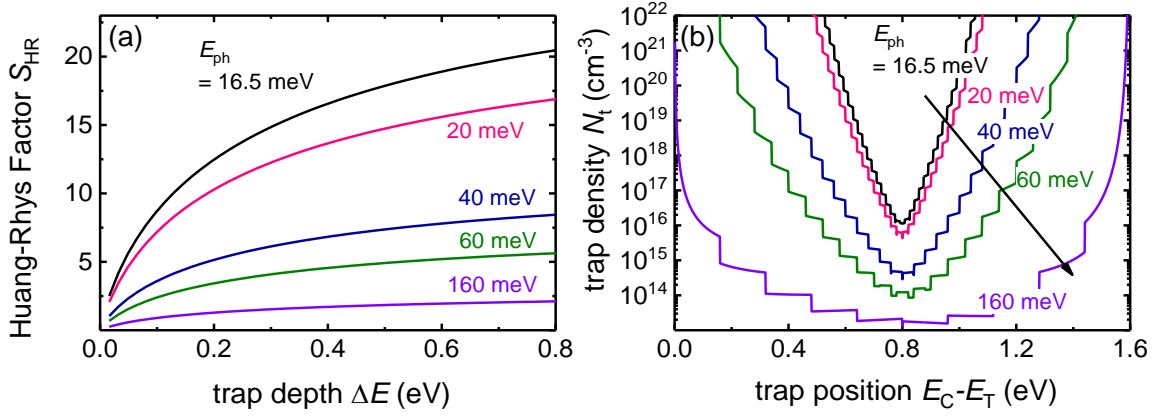

**Figure 3:** (a) Huang-Rhys factor for a positively charged trap as a function of $\Delta E$ and the phonon energy. All parameters were kept as defined in Table I, however, the product $M_r E_{ph}$ was kept constant (see SI). (b) Trap density needed to achieve a SRH lifetime of 1 µs, for phonon energies ranging from 16.5 meV to 160 meV, showing that larger phonon energies make the exact position of the trap less important with midgap energy always remaining the worst-case scenario.

**Table I:** Parameter and abbreviations used in the simulations.

| Parameter | Symbol or Equation | Value |
|---|---|---|
| phonon energy | $E_{ph} = \hbar\omega$ | 16.5 meV (LO Phonon)[45] |
| reduced mass | $M_r = M_{Pb}M_I/(M_{Pb}+M_I)$ | 78.7 |
| permittivity(frequency) | $\varepsilon(f)$ | $\varepsilon(0) = 33.5\varepsilon_0$ |
| | | $\varepsilon(\infty) = 5.0\varepsilon_0$[45] |
| lattice constant | $a_0$ | 6.3 Å[84] |
| radius of sphere with Brillouin zone volume | $q_D = (6\pi^2)^{1/3}/a_0$ | 6.2 nm$^{-1}$ |
| effective mass | $m_{eff}$ | 0.2 (assumed equal for electrons and holes) |
| Bohr radius | $a_H = 4\pi\varepsilon_0\hbar^2/mq^2$ | $5.292\times10^{-2}$ nm |
| effective Bohr radius | $a^* = a_H\, \varepsilon_r(0)/m_{eff}$ | 8.9 nm |
| Rydberg energy | $R_H = q^2/(8\pi\varepsilon_0 a_H)$ | 13.605 eV |
| effective Rydberg energy | $R^* = q^2/(8\pi\varepsilon(0)a^*)$ | 2.4 meV |
| | $\nu = q(8\pi\varepsilon(0)a^*\Delta E)^{-1/2}$ | variable |
| Sommerfeld factor[85] | $s_a = 4(\pi R^*/kT)^{1/2}$ | 2.2 |